\documentclass[10pt,conference]{IEEEtran}
\usepackage{graphicx} 
\usepackage{epsfig}
\usepackage{cite}
\usepackage{times}
\usepackage{comment}
\usepackage{amsmath}
\usepackage{xcolor}
\usepackage{soul,color}
\usepackage{lipsum}  
\usepackage{tikz}
\usepackage{pgfplots}
\pgfplotsset{compat=1.18}
\usepackage{booktabs}
\usepackage{array}
\usepackage{tabularx}
\usepackage{csquotes}
\usepackage{gensymb}  
\usepackage{textcomp}
\usepackage{amsmath,amssymb,amsfonts}
\usepackage{mathtools} 
\usepackage{epstopdf}
 \usepackage[nolist]{acronym}
\usepackage{tabularx}

\title{EMF-aware Radcom}
\author{m.chemingui }
\date{December 2023}

\title{EMF-Aware Waveform for Dual-functional Radar Communication Systems}
\author{
    Mariem Chemingui, 
    Ahmed Elzanaty,
    Rahim Tafazolli \\
    Institute for Communication Systems (ICS), University of Surrey, UK,\\ 
   (m.chemingui, a.elzanaty, r.tafazolli)@surrey.ac.uk

\thanks{This work is supported by the UK Department for Science, Innovation and Technology under the Future Open Networks Research Challenge project TUDOR (Towards Ubiquitous 3D Open Resilient Network). The views expressed are those of the authors and do not necessarily represent the project.}
}
\IEEEoverridecommandlockouts\IEEEpubid{\makebox[\columnwidth]{} \hspace{\columnsep}\makebox[\columnwidth]{ }}
 \IEEEoverridecommandlockouts

\begin{document}
\begin{acronym}
\acro{5G-NR}{5G New Radio}
\acro{3GPP}{3rd Generation Partnership Project}
\acro{AC}{address coding}
\acro{ACF}{autocorrelation function}
\acro{ACR}{autocorrelation receiver}
\acro{ADC}{analog-to-digital converter}
\acrodef{aic}[AIC]{Analog-to-Information Converter}     
\acro{AIC}[AIC]{Akaike information criterion}
\acro{aric}[ARIC]{asymmetric restricted isometry constant}
\acro{arip}[ARIP]{asymmetric restricted isometry property}
\acro{ARIS}{Aerial RIS}

\acro{ARQ}{automatic repeat request}
\acro{AUB}{asymptotic union bound}
\acrodef{awgn}[AWGN]{Additive White Gaussian Noise}     
\acro{AWGN}{additive white Gaussian noise}
\acro{APSK}[PSK]{asymmetric PSK} 
\acro{AO}{alternate optimization}
\acro{IoT}{Internet of Things}
\acro{IoE}{Internet of Everything}
\acro{ZF}{Zero-Forcing}

\acro{mmW}{Millimeter Waves}

\acro{waric}[AWRICs]{asymmetric weak restricted isometry constants}
\acro{warip}[AWRIP]{asymmetric weak restricted isometry property}
\acro{MUI}{multi-user interference}
\acro{BCH}{Bose, Chaudhuri, and Hocquenghem}        
\acro{BCHC}[BCHSC]{BCH based source coding}
\acro{BEP}{bit error probability}
\acro{BFC}{block fading channel}
\acro{BG}[BG]{Bernoulli-Gaussian}
\acro{BGG}{Bernoulli-Generalized Gaussian}
\acro{BPAM}{binary pulse amplitude modulation}
\acro{BPDN}{Basis Pursuit Denoising}
\acro{BPPM}{binary pulse position modulation}
\acro{BPSK}{binary phase shift keying}
\acro{BPZF}{bandpass zonal filter}
\acro{BSC}{binary symmetric channels}              
\acro{BU}[BU]{Bernoulli-uniform}
\acro{BER}{bit error rate}
\acro{BS}{base station}
\acro{EI}{exposure index}
\acro{MU-MIMO}{multi-user multiple-input-multiple-output}
\acro{MU-SIMO}{multi-user single-input-multiple-output}

\acro{CP}{Cyclic Prefix}
\acrodef{cdf}[CDF]{cumulative distribution function}   
\acro{CDF}{cumulative distribution function}
\acrodef{c.d.f.}[CDF]{cumulative distribution function}
\acro{CCDF}{complementary cumulative distribution function}
\acrodef{ccdf}[CCDF]{complementary CDF}       
\acrodef{c.c.d.f.}[CCDF]{complementary cumulative distribution function}
\acro{CD}{cooperative diversity}
\acro{CDMA}{Code Division Multiple Access}
\acro{ch.f.}{characteristic function}
\acro{CIR}{channel impulse response}
\acro{cosamp}[CoSaMP]{compressive sampling matching pursuit}
\acro{CR}{cognitive radio}
\acro{cs}[CS]{compressed sensing}                  
\acrodef{cscapital}[CS]{Compressed sensing} 
\acrodef{CS}[CS]{compressed sensing}
\acro{CSI}{channel state information}
\acro{CCSDS}{consultative committee for space data systems}
\acro{CC}{convolutional coding}
\acro{Covid19}[COVID-19]{Coronavirus disease}

\acro{DAA}{detect and avoid}
\acro{DAB}{digital audio broadcasting}
\acro{DCT}{discrete cosine transform}
\acro{dft}[DFT]{discrete Fourier transform}
\acro{DFRC}{dual-functional radar communication}
\acro{DR}{distortion-rate}
\acro{DS}{direct sequence}
\acro{DS-SS}{direct-sequence spread-spectrum}
\acro{DTR}{differential transmitted-reference}
\acro{DVB-H}{digital video broadcasting\,--\,handheld}
\acro{DVB-T}{digital video broadcasting\,--\,terrestrial}
\acrodef{DL}{downlink}
\acro{DSSS}{Direct Sequence Spread Spectrum}
\acro{DFT-s-OFDM}{Discrete Fourier Transform-spread-Orthogonal Frequency Division Multiplexing}
\acro{DAS}{distributed antenna system}
\acro{DNA}{Deoxyribonucleic Acid}

\acro{EC}{European Commission}
\acro{EED}[EED]{exact eigenvalues distribution}
\acro{EIRP}{Equivalent Isotropically Radiated Power}
\acro{ELP}{equivalent low-pass}
\acro{eMBB}{Enhanced Mobile Broadband}
\acro{EMF}{electromagnetic field}
\acro{EM}{electromagnetic}
\acro{EU}{European union}

\acro{FC}[FC]{fusion center}
\acro{FCC}{Federal Communications Commission}
\acro{FEC}{forward error correction}
\acro{FFT}{fast Fourier transform}
\acro{FH}{frequency-hopping}
\acro{FH-SS}{frequency-hopping spread-spectrum}
\acrodef{FS}{Frame synchronization}
\acro{FSsmall}[FS]{frame synchronization}  
\acro{FDMA}{Frequency Division Multiple Access}

\acro{GA}{Gaussian approximation}
\acro{GF}{Galois field }
\acro{GG}{Generalized-Gaussian}
\acro{GIC}[GIC]{generalized information criterion}
\acro{GLRT}{generalized likelihood ratio test}
\acro{GPS}{Global Positioning System}
\acro{GMSK}{Gaussian minimum shift keying}
\acro{GSMA}{Global System for Mobile communications Association}
\acro{GR}{Gaussian randomization}
\acro{HAP}{high altitude platform}

\acro{IDR}{information distortion-rate}
\acro{IFFT}{inverse fast Fourier transform}
\acro{iht}[IHT]{iterative hard thresholding}
\acro{i.i.d.}{independent, identically distributed}
\acro{IoT}{Internet of Things}                      
\acro{IR}{impulse radio}
\acro{lric}[LRIC]{lower restricted isometry constant}
\acro{lrict}[LRICt]{lower restricted isometry constant threshold}
\acro{ISI}{intersymbol interference}
\acro{ITU}{International Telecommunication Union}
\acro{ICNIRP}{International Commission on Non-Ionizing Radiation Protection}
\acro{IEEE}{Institute of Electrical and Electronics Engineers}
\acro{ICES}{IEEE international committee on electromagnetic safety}
\acro{IEC}{International Electrotechnical Commission}
\acro{IARC}{International Agency on Research on Cancer}
\acro{IS-95}{Interim Standard 95}

\acro{LEO}{low earth orbit}
\acro{LF}{likelihood function}
\acro{LLF}{log-likelihood function}
\acro{LLR}{log-likelihood ratio}
\acro{LLRT}{log-likelihood ratio test}
\acro{LOS}{Line-of-Sight}
\acro{LRT}{likelihood ratio test}
\acro{wlric}[LWRIC]{lower weak restricted isometry constant}
\acro{wlrict}[LWRICt]{LWRIC threshold}
\acro{LPWAN}{low power wide area network}
\acro{LoRaWAN}{Low power long Range Wide Area Network}
\acro{NLOS}{non-line-of-sight}

\acro{MB}{multiband}
\acro{MC}{multicarrier}
\acro{MDS}{mixed distributed source}
\acro{MF}{matched filter}
\acro{m.g.f.}{moment generating function}
\acro{MI}{mutual information}
\acro{MIMO}{multiple-input multiple-output}
\acro{MISO}{multiple-input single-output}
\acrodef{maxs}[MJSO]{maximum joint support cardinality}                       
\acro{ML}[ML]{maximum likelihood}
\acro{MMSE}{minimum mean-square error}
\acro{MMV}{multiple measurement vectors}
\acrodef{MOS}{model order selection}
\acro{M-PSK}[${M}$-PSK]{$M$-ary phase shift keying}                       
\acro{M-APSK}[${M}$-PSK]{$M$-ary asymmetric PSK} 

\acro{M-QAM}[$M$-QAM]{$M$-ary quadrature amplitude modulation}
\acro{MRC}{maximal ratio combiner}                  
\acro{maxs}[MSO]{maximum sparsity order}                                      
\acro{M2M}{machine to machine}                                                
\acro{MUI}{multi-user interference}
\acro{mMTC}{massive Machine Type Communications}      
\acro{mm-Wave}{millimeter-wave}
\acro{MP}{mobile phone}
\acro{MPE}{maximum permissible exposure}
\acro{MAC}{media access control}
\acro{MUMIMO}{multi-user  \ac{MIMO}}

\acro{NLoS}{non line-of-sight}
\acro{NB}{narrowband}
\acro{NBI}{narrowband interference}
\acro{NLA}{nonlinear sparse approximation}
\acro{NTIA}{National Telecommunications and Information Administration}
\acro{NTP}{National Toxicology Program}
\acro{NHS}{National Health Service}

\acro{OC}{optimum combining}                       \acro{QCQP}  {quadratically constrained quadratic problem}
\acro{OC}{optimum combining}
\acro{ODE}{operational distortion-energy}
\acro{ODR}{operational distortion-rate}
\acro{OFDM}{orthogonal frequency-division multiplexing}
\acro{omp}[OMP]{orthogonal matching pursuit}
\acro{OSMP}[OSMP]{orthogonal subspace matching pursuit}
\acro{OQAM}{offset quadrature amplitude modulation}
\acro{OQPSK}{offset QPSK}
\acro{OFDMA}{Orthogonal Frequency-division Multiple Access}
\acro{OPEX}{Operating Expenditures}
\acro{OQPSK/PM}{OQPSK with phase modulation}

\acro{PAM}{pulse amplitude modulation}
\acro{PAR}{peak-to-average ratio}
\acrodef{pdf}[PDF]{probability density function}                      
\acro{PDF}{probability density function}
\acrodef{p.d.f.}[PDF]{probability distribution function}
\acro{PDP}{power dispersion profile}
\acro{PMF}{probability mass function}                             
\acrodef{p.m.f.}[PMF]{probability mass function}
\acro{PN}{pseudo-noise}
\acro{PPM}{pulse position modulation}
\acro{PRake}{Partial Rake}
\acro{PSD}{power spectral density}
\acro{PSEP}{pairwise synchronization error probability}
\acro{PSK}{phase shift keying}
\acro{PD}{power density}
\acro{8-PSK}[$8$-PSK]{$8$-phase shift keying}

\acro{FSK}{frequency shift keying}

\acro{QAM}{Quadrature Amplitude Modulation}
\acro{QPSK}{quadrature phase shift keying}
\acro{OQPSK/PM}{OQPSK with phase modulator }
\acro{RadCom}{dual-functional radar communication}
\acro{RE}{Resource Element}
\acro{RD}[RD]{raw data}
\acro{RDL}{"random data limit"}
\acro{ric}[RIC]{restricted isometry constant}
\acro{rict}[RICt]{restricted isometry constant threshold}
\acro{rip}[RIP]{restricted isometry property}
\acro{ROC}{receiver operating characteristic}
\acro{rq}[RQ]{Raleigh quotient}
\acro{RS}[RS]{Reed-Solomon}
\acro{RSC}[RSSC]{RS based source coding}
\acro{r.v.}{random variable}                               
\acro{R.V.}{random vector}
\acro{RMS}{root mean square}
\acro{RFR}{radiofrequency radiation}
\acro{RIS}{reconfigurable intelligent surface}
\acro{RNA}{RiboNucleic Acid}

\acro{LoS}{line-of-sight}

\acro{SA}[SA-Music]{subspace-augmented MUSIC with OSMP}
\acro{SAR}{specific absorption rate}
\acro{SCBSES}[SCBSES]{Source Compression Based Syndrome Encoding Scheme}
\acro{SCM}{sample covariance matrix}
\acro{SDR}{semi-definite relaxation}
\acro{SDP}{semi-definite program}
\acro{SEP}{symbol error probability}
\acro{SG}[SG]{sparse-land Gaussian model}
\acro{SIMO}{single-input multiple-output}
\acro{SINR}{signal-to-interference plus noise ratio}
\acro{SIR}{signal-to-interference ratio}
\acro{SISO}{single-input single-output}
\acro{SMV}{single measurement vector}
\acro{SNR}[\textrm{SNR}]{signal-to-noise ratio} 
\acro{sp}[SP]{subspace pursuit}
\acro{SS}{spread spectrum}
\acro{SW}{sync word}
\acro{SAR}{specific absorption rate}
\acro{SSB}{synchronization signal block}
\acro{SCA}{successive convex approximation}
\acro{SER}{symbol error rate }

\acro{TH}{time-hopping}
\acro{ToA}{time-of-arrival}
\acro{TR}{transmitted-reference}
\acro{TW}{Tracy-Widom}
\acro{TWDT}{TW Distribution Tail}
\acro{TCM}{trellis coded modulation}
\acro{TDD}{time-division duplexing}
\acro{TDMA}{Time Division Multiple Access}

\acro{UAV}{unmanned aerial vehicle}
\acro{uric}[URIC]{upper restricted isometry constant}
\acro{urict}[URICt]{upper restricted isometry constant threshold}
\acro{UWB}{ultrawide band}
\acro{UWBcap}[UWB]{Ultrawide band}   
\acro{URLLC}{Ultra Reliable Low Latency Communications}
         
\acro{wuric}[UWRIC]{upper weak restricted isometry constant}
\acro{wurict}[UWRICt]{UWRIC threshold}                
\acro{UE}{user equipment}
\acrodef{UL}{uplink}

\acro{WiM}[WiM]{weigh-in-motion}
\acro{WLAN}{wireless local area network}
\acro{wm}[WM]{Wishart matrix}                               
\acroplural{wm}[WM]{Wishart matrices}
\acro{WMAN}{wireless metropolitan area network}
\acro{WPAN}{wireless personal area network}
\acro{wric}[WRIC]{weak restricted isometry constant}
\acro{wrict}[WRICt]{weak restricted isometry constant thresholds}
\acro{wrip}[WRIP]{weak restricted isometry property}
\acro{WSN}{wireless sensor network}                        
\acro{WSS}{wide-sense stationary}
\acro{WHO}{World Health Organization}
\acro{Wi-Fi}{wireless fidelity}

\acro{sss}[SpaSoSEnc]{sparse source syndrome encoding}

\acro{VLC}{visible light communication}
\acro{VPN}{virtual private network} 
\acro{RF}{radio frequency}
\acro{FSO}{free space optics}
\acro{IoST}{Internet of space things}

\acro{GSM}{Global System for Mobile Communications}
\acro{2G}{second-generation cellular network}
\acro{3G}{third-generation cellular network}
\acro{4G}{fourth-generation cellular network}
\acro{5G}{5th-generation}	
\acro{gNB}{next generation node B base station}
\acro{NR}{New Radio}
\acro{UMTS}{Universal Mobile Telecommunications Service}
\acro{LTE}{Long Term Evolution}

\acro{QoS}{quality of service}
\acro{KKT}{Karush–Kuhn–Tucker}

\acro{WPD}{wireless power transfer device}
\acro{w.r.t}{with respect to}
\end{acronym}

\maketitle
 \IEEEpubid{{Accepted for presentation at IEEE PIMRC 2024. \hfill}}
\begin{abstract}
Emerging \ac{RadCom} systems promise to revolutionize wireless systems by enabling radar sensing and communication on a shared platform, thereby enhancing spectral efficiency. However, the high transmit power required for efficient radar operation poses risks by potentially exceeding the \ac{EMF} exposure limits enforced by the regulations. To address this challenge, we propose an \ac{EMF}-aware signalling design that enhances \ac{RadCom} system performance while complying with \ac{EMF} constraints. Our approach considers exposure levels not only experienced by network users but also in sensitive areas such as schools and hospitals, where the exposure must be further reduced.
First, we model the exposure metric for the users and the sectors that encounter sensitive areas. Then, we design the waveform by exploiting the trade-off between radar and communication while satisfying the exposure constraints. We reformulate the problem as a convex optimization program and solve it in closed form using \ac{KKT} conditions. The numerical results demonstrate the feasibility of developing a robust \ac{RadCom} system with low \ac{EM} radiations.
\end{abstract}
\pagenumbering{gobble}
\section{Introduction} 
With the introduction of numerous connected devices, spectrum congestion has become a real challenge in communication systems \cite{SpectrumCongestion}. In this regard, several studies have recently investigated \ac{RadCom} systems using the same time-frequency resources to enhance spectrum utilization. Specifically, radar sensing and communications could be jointly performed by the \ac{BS} to efficiently utilize the frequency resources \cite{8386661,7485158}. Performing radar sensing and
communication on the same frequency range will also substantially reduce the energy consumption compared to a system where both operations are performed on two dedicated platforms \cite{8828023,9226446}.

One approach to realize \ac{RadCom} is to conduct a signalling design, occasionally referred as waveform design, that embeds communication symbols for multiple users in the radar beampattern. For instance, in \cite{7485158}, the signal was synthesized such that the  \ac{SINR} of the radar is maximized while satisfying a constraint on the sum data rate.  Moreover, the authors in \cite{8454491}, considered enhancing energy efficiency in \ac{RadCom} systems, while guaranteeing the required probability of detection for radar targets and the required data rate for users. In \cite{7962141}, the radar waveform was designed to maximize the probability of detection while satisfying the information rate for the communication receivers.

Nevertheless, prior studies have not considered the \ac{EMF} exposure resulting from the high transmit power typically required for efficient radar operation \cite{RadarEMF}.
In this context, prioritizing the safety of network users and the inhabitants in sensitive areas becomes paramount. In fact, the radio-frequency exposure has been classified as \textquote{Possibly carcinogenic to humans} (Group 2B) \cite{iarc2013non}. To address these concerns, regulatory bodies such as the \ac{ICNIRP} have established guidelines specifying that the maximum user exposure to \ac{EM} radiations should remain below certain thresholds \cite{ElzanatySurvey}. Each country establishes its own safety regulations concerning \ac{EMF}, primarily guided by existing protocols \cite{world2017exposure}. In addition to these guidelines, certain countries implement further precautionary measures, such as creating restricted/sensitive areas, e.g., schools and hospitals, where the exposure level is very low compared to other areas \cite{ElzanatySurvey,AEStochasticGeo}.

In this regard, \ac{EMF}-aware techniques have been investigated in several communication scenarios, e.g., \ac{MIMO} and \ac{RIS}-aided networks \cite{FabienOFDM,10167745}. Nevertheless, an \ac{EMF}-aware design for \ac{RadCom} has not been investigated, albeit the importance of meeting regulatory requirements and analyzing the impact of limiting the \ac{EMF} exposure in sensitive areas on the system performance.

To this end, we propose a novel framework to analyze the impact of \ac{EMF} exposure constraints in \ac{RadCom}. More precisely, the proposed scheme considers the restrictions on the \ac{EMF} exposure for the network users and within restricted areas, while achieving a trade-off between the radar and communication functionalities. The main contributions of this paper can be summarized as follows:
\begin{itemize}
    \item We analyse the \ac{EMF} exposure in \ac{RadCom} systems.
    \item We propose a dual-functional signalling design while limiting the \ac{EMF} exposure.
     \item We investigate the trade-off between the radar, communication, and \ac{EMF} exposure compliance.
    \item We propose a Pareto optimization problem to minimize the multi-user interference and the error between the desired and designed radar waveform with \ac{EMF} constraints.
     \vspace{400px}
   \IEEEpubidadjcol
    \item We reformulate the problem as a convex optimization problem and provide a closed-form expression for the waveform.
\end{itemize}
Throughout this work, lowercase letters (e.g., $x$), bold lowercase letters (e.g., $\boldsymbol{x}$), and bold uppercase letters (e.g., $\boldsymbol{X}$) denote scalars, vectors, and matrices, respectively.
We use $\mathbb{R}$ to denote the set of real numbers and $\mathbb{C}$ denote the set of complex numbers, where $j$ represents the imaginary unit.
For a matrix $\boldsymbol{A}$, $\boldsymbol{A}^{\mathrm{T}}$ refers to its transpose, and $\boldsymbol{A}^{\mathrm{H}}$ refers to its conjugate transpose. The operator $\mathrm{Tr}(\cdot)$ denotes the trace of a matrix, the operator $\mathrm{Re}\left\{\cdot\right\}$ represents the real part of a complex entity, and the operator $\|\cdot\|_{\text{F}}$ represents the Frobenius norm. The optimal solution is denoted by $(.)^*$.
\section{System Model and Problem Formulation}
\label{Sec2}
This section describes the system model and problem formulation for the considered \ac{RadCom} system with \ac{EMF} constraints.
\subsection{Channel Model}
\begin{figure}[t]
    \centering
\includegraphics[width=0.9\linewidth]{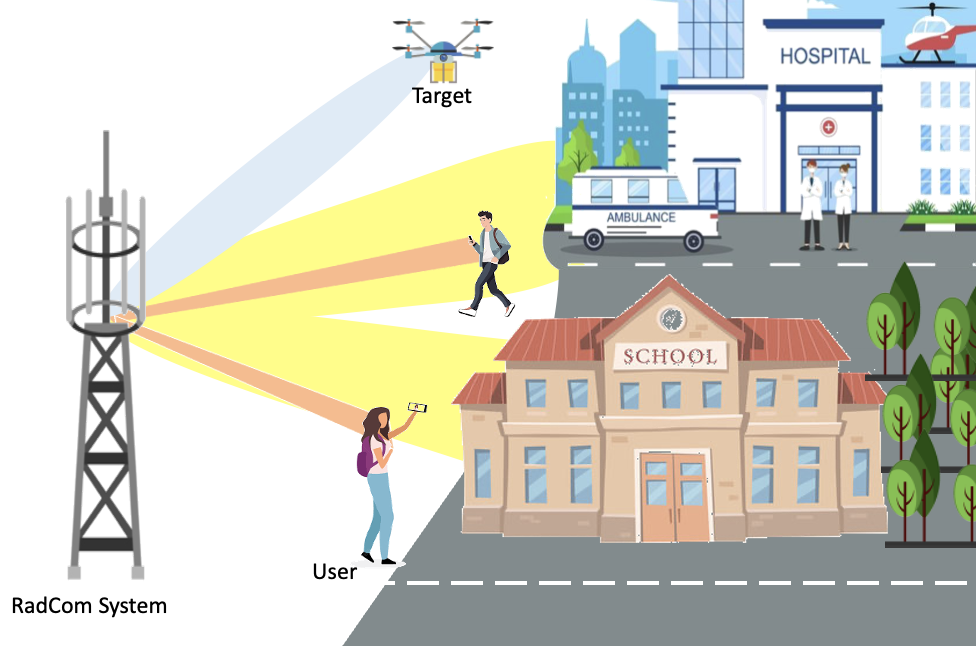}
    \caption{Illustration of RadCom network with restricted areas.}
    \label{fig:sys}
\end{figure}
We consider a \ac{RadCom} system transmitting radar waveforms to the targets, which also embeds the users' communication symbols.
The \ac{BS} is equipped with an array of $N$ antennas where the antenna reference is positioned at $\mathbf{q}_{\text{B}} = [x_{\text{B}}, y_{\text{B}}, z_{\text{B}}]^{\text{T}}$. The \ac{BS} serves $U$ single-antenna users located at $\mathbf{q}_u = [x_u, y_u, 0]^{\text{T}}, \forall u \in \mathcal{U} = \{1,\cdots,U\},$ while detecting radar targets at the same time, as shown in Fig.~\ref{fig:sys}. Also, we consider $A$ sensitive areas where the exposure should be maintained below a predefined limit, usually stricter than the typical regulation thresholds.

For users in the far-field, the distance between the $u^{\text{th}}$ user and the $n^{\text{th}}$ antenna can be written as 
\begin{multline}
        d_{n,u} = (n-1)~\delta~ \cos \theta_u + d_u,\\ \forall n \in \{1,2,\cdots,N\}, \forall u \in  \mathcal{U},
\end{multline}
where $\delta$ denotes the antenna spacing,
$\theta_u$ and $d_u = \|\mathbf{q}_{\text{B}}-\mathbf{q}_u \|$ are the angle and the distance between the user $u$ and the reference antenna at the \ac{BS}, respectively. We consider a geometric model for the channel between the \ac{BS} and users. Let $\mathbf{H} =[\mathbf{h}_1,\mathbf{h}_2,\dots,\mathbf{h}_U]^{\text{T}} \in \mathbb{C}^{U\times N}$ be the channel matrix, where 
\begin{align}
    \mathbf{h}_u \triangleq \beta_u ~ e^{-j\phi_u}~ \mathbf{a}(\theta_u),\quad \forall u \in \mathcal{U},
\end{align}
is the communication channel between the \ac{BS} and the $u^{\text{th}}$ user, $\beta_u = \frac{\lambda}{4\pi d_u} e^{-j\frac{2\pi}{\lambda}d_u} $

is the signal attenuation coefficient depending on the distance between the \ac{BS} and the user $u$ where $\lambda$ denotes the wavelength, and $\phi_u$ is a uniformly distributed random variable between $0$ and $2\pi$ representing the channel phase at the reference antenna. The steering vector $\mathbf{a}$ is defined as follows
\begin{equation}
\mathbf{a}(\theta) \triangleq \frac{1}{\sqrt{N}}\left[1,e^{-j k \cos \theta},\ldots,e^{-j (N-1) k \cos\theta}\right]^{\text{T}},
\end{equation}
where $k={2\pi d}/{\lambda}$ is the wave vector.
\subsection{Signal Model}
The \ac{BS} transmits the downlink signal $\mathbf{X}=[\mathbf{x}_1,\mathbf{x}_2\dots,\mathbf{x}_L] \in \mathbb{C}^{N\times L}$, where $L$ is the length of the communication frame.
The received symbol matrix at the downlink users can be expressed as 
\begin{equation} 
\mathbf {Y}  = \mathbf{H}\mathbf{X} + \mathbf {W}, \tag{1} \end{equation}
where each element of the noise matrix $\mathbf{W}$ is a complex Gaussian random variable with variance $\sigma_0^2$. Consequently, for a given constellation matrix $\mathbf{S} \in \mathbb{C}^{U \times L}$ holding the information symbols for the users at different time instants, the received signal can be rewritten as  
\begin{equation}
    \mathbf{Y} = \mathbf{S}+(\mathbf{H}\mathbf{X} - \mathbf{S}) + \mathbf{W},
    \label{eq.receivedsignal}
\end{equation}
where the term $(\mathbf{H} \mathbf{X} - \mathbf{S})$ represents the \ac{MUI} signals with power equals to $\left\Vert \mathbf {H}\mathbf {X} - {\mathbf {S}} \right\Vert _F^2
$, which is considered as the performance metric for the communication part of our system. 

The radiation beampattern of the \ac{BS} can be expressed as 
\begin{equation}
    \mathrm{P}(\theta) = \frac{1}{4\pi}\mathbf{a}^{\text{H}}(\theta)~\mathbf{R}~\mathbf{a}(\theta), \quad\forall \theta \in \left[0,2\pi\right],
    \label{beampattern}
\end{equation}
 where $\mathbf{R} \triangleq \mathbf{X}\mathbf{X}^{\mathrm{H}}$ is the empirical covariance matrix of the transmitted signal.
 
 Equations \eqref{eq.receivedsignal} and \eqref{beampattern} indicate that both communication and radar performances depend on the transmitted waveform  $\mathbf{X}$ either directly or through its covariance matrix.
 
\subsection{Downlink EMF Exposure Model}
The \ac{EMF}-aware design is especially crucial in environments where significant radiated power is required for effective system operation, such as detecting and tracking targets while ensuring effective communication. To address this, we assess the downlink \ac{EMF} radiation in terms of the incident power density metric which quantifies the power radiated per unit surface at a given distance \cite{icnirp2020,7032050}. This metric is also proportional to the whole-body average \ac{SAR} \cite{icnirp2020}. 

We consider analyzing the incident power density for two entities: \textit{(i)} the users; \textit{(ii)} the sensitive areas. Regarding the \ac{EMF} exposure of the users, the incident power density at the $u^\text{th}$ user can be written as 
  \begin{equation}
      \mathrm{S}_{\text{U}}(\theta_{u},d_{u}) = \frac{1}{4\pi d_u^2}~\mathbf{a}^{\textrm{H}}(\theta_u) ~ \mathbf{R} ~\mathbf{a}(\theta_u)~[\text{W~m}^{-2}].
      \label{radiation}
  \end{equation}
For the exposure in sensitive areas, we consider the average power density over the area. Let $\mathcal{A}=\{1,\dots,a,\dots,A\}$ denote the set of restricted areas, where each sector $a$ is defined by its angular domain represented by the initial angle $\theta_{\text{I}_a}$, the final angle $\theta_{\text{F}_a}$, and its closest distance from the associated \ac{BS} $d_a$. Each sector is considered as a segment of a circle, bounded by the specified angles. More precisely, the sector is confined within an arc created by $\theta_{\text{I}_a}$ and $\theta_{\text{F}_a}$, where $d_a$ is the distance between the \ac{BS} and the arc. Therefore, the average power incident density over sector $a$ can be computed as
\begin{equation}
   \mathrm{S}_{\text{A}}(D_a)=\frac{1}{4\pi d_a^2  \Delta_a} \!\int_{\theta_{\text{I}_a}}^{\theta_{\text{F}_a}} \!\mathbf{a}^{\textrm{H}}(\theta) ~ \mathbf{R} ~\mathbf{a}(\theta) ~\sin(\theta)~d\theta ~[\text{W m}^{-2}],
\end{equation}
where $D_a = (\theta_{\text{I}_a},\theta_{\text{F}_a},d_a)$ denotes the sector $a$ parameters and $\Delta_a =|\theta_{\text{F}_a}-\theta_{\text{I}_a}|$ denotes its angular domain. 

Similar to what we noticed in \eqref{eq.receivedsignal} and \eqref{beampattern}, the covariance matrix of the signal impacts also the exposure in terms of power density, playing a significant role in determining the \ac{EM} radiations.

\section{Problem Formulation}
\label{Sec3}
In this section, we formulate an optimization problem for \ac{EMF}-aware dual-functional radar communication signalling design. To begin, we have to determine the waveform covariance matrix $\mathbf{R}_0$ that is optimized for radar performance, i.e., closest to the desired radar beampattern, guaranteeing that each antenna element transmits equal power, and the total transmitted power is $P$. Hence, $\mathbf{R}_0$ can be obtained by minimizing the error between the desired and optimized beampatterns, as outlined in \cite{RadComUsingCrossCorrelation}. Then, for a given $\mathbf{R}_0$, we need to synthesize the desired waveform $\mathbf{X}_0$ according to additional system requirements \cite{RadComUsingCrossCorrelation}. In our study, we specifically design the directional waveform $\mathbf{X}_0$ by minimizing the \ac{MUI}, i.e.,
\begin{subequations}
\begin{align}
\mathbf{X}_0 =\underset{\mathbf{X}  \in \mathbb{C}^{N\times L}}{\text{argmin}} \quad &  \left\Vert \mathbf {H}\mathbf {X} - {\mathbf {S}} \right\Vert _{\text{F}}^2  \\
\text{subject to} \quad & \mathbf{X}\mathbf{X}^{\text{H}} = \mathbf{R}_{0}.
\label{X0}
\end{align}
\end{subequations}
The waveform $\mathbf{X}_0$ is designed to form multiple transmit beams in the direction of targets of interest. However, this synthesis is limited to radar purposes. Therefore, to introduce more flexibility in the design of a suitable waveform for joint radar and communication, we exploit the trade-off between the two functionalities. To this end, we formulate the optimization problem considering the minimization of \ac{MUI} and the discrepancy between the required and optimized radar waveforms while limiting the downlink exposure levels as follows
\begin{subequations}
\begin{align}
\underset{\mathbf{X}}{\text{minimize}}  \quad & \rho \left\Vert \mathbf {H}\mathbf {X} - {\mathbf {S}} \right\Vert _{\text{F}}^2 + (1-\rho) \left\Vert \mathbf {X} - {\mathbf {X}_0} \right\Vert _{\text{F}}^2  \\
\text{subject to}\quad & \left\Vert \mathbf{X} \right\Vert _{\text{F}}^2 \leq P_{\mathrm{max}},  \label{powConst}\\
& \mathrm{S}_{\text{U}}(\theta_u, d_u) \leq \Gamma_u, \quad \forall u \in \mathcal{U}, \label{UExp} \\
& \mathrm{S}_{\text{A}}(D_a) \leq \Gamma_a, \quad \quad~\forall a \in \mathcal{A} \label{SExp},
\end{align}
\label{optPb1}
\end{subequations}
where $\rho \in \left[0,1\right]$ denotes the trade-off factor indicating the priority of the two performance metrics, $P_{\text{max}}$ is the maximum transmit power, $\Gamma_u$ and $\Gamma_a$ denote the maximum allowed power densities for the users and sensitive areas, respectively. 
The first constraint \eqref{powConst} refers to the control of power, and the second constraint \eqref{UExp} is the exposure constraint where the received power density should not be greater than the threshold for all the users. The third constraint \eqref{SExp} refers to the sectors where the downlink exposure should be further restricted. 
%
The problem \eqref{optPb1} can be rewritten in a quadratic form as 
\begin{subequations}
\begin{align}
\underset{\mathbf{X}}{\text{minimize}} \quad &  \mathrm{Tr}(\mathbf{X}^{\mathrm{H}}\mathbf{A}\mathbf{X}) + 2~\mathrm{Re}\left\{\mathrm{Tr}(\mathbf{X}^{\mathrm{H}}\mathbf{B})\right\} \\
\text{subject to} \quad & \left\Vert \mathbf{X} \right\Vert_{\text{F}}^2 \leq P_{\mathrm{max}}, \label{powConst1}\\
& \frac{1}{4\pi d_u^2} \left\Vert \bar{\mathbf{a}}_u^{\mathrm{H}}~\mathbf{X} \right\Vert_{\text{F}} ^2 \leq \Gamma_u,\quad \forall u \in \mathcal{U},\label{UExp1}\\
&  \frac{1}{4\pi d_a^2} \left\Vert \bar{\mathbf{b}}_a^{\mathrm{H}}~
 \mathbf{X} \right\Vert_{\text{F}} ^2 \leq \Gamma_a,\quad \forall a \in \mathcal{A}, \label{SExp1}
\end{align}
\end{subequations}
where $\mathbf{A} \triangleq \rho~\mathbf{H}^{\mathrm{H}}\mathbf{H}+(1-\rho )\mathbf{I}_N $ and $\mathbf{B} \triangleq \rho ~\mathbf{H}^{\mathrm{H}} \mathbf{A}+(1-\rho) \mathbf{X}_0$. For the sake of simplicity, the explicit representation of the angles is omitted. Let $\bar{\mathbf{a}}_u =\mathbf{a}(\theta_u)$ and $\bar{\mathbf{b}}_a = \left(\mathbf{b}^{\mathrm{H}}(\theta_{\text{F}_a})-\mathbf{b}^{\mathrm{H}}(\theta_{\text{I}_a})\right)$, where vector
$\mathbf{b}$ is expressed as 
\begin{equation}
    \mathbf{b}(\theta) \triangleq \frac{-1}{\sqrt{N}}\left[\cos\theta, \frac{j e^{-j k \cos \theta}}{k}\!,\ldots,\!\frac{j e^{-jk (N-1) \cos\theta}}{k (N-1)}\right]^{\text{T}}\!.
\end{equation}

The convex nature of the reformulated problem allows us to apply the \ac{KKT} optimality conditions, which are both necessary and sufficient for optimal solutions. Let us define the Lagrangian function of the reformulated \ac{EMF}-aware problem as follows 
\begin{equation}
\begin{aligned}
&\mathcal{L}(\mathbf{X}, \alpha, \boldsymbol{\gamma}) = \mathrm{Tr}(\mathbf{X}^{\text{H}}\mathbf{A}\mathbf{X}) + 2~\mathrm{Re}\left\{\mathrm{Tr}(\mathbf{X}^{\mathrm{H}}\mathbf{B})\right\} \\
& + \alpha\!\left(\left\Vert\mathbf{X} \right\Vert_{\text{F}} ^2 - P_{\text{max}}\right) + \sum_{u=1}^{U}\gamma_u\!\left(\frac{1}{4\pi d_u^2} \!\left\Vert \bar{\mathbf{a}}_u^\mathrm{H}~\mathbf{X} \right\Vert_{\text{F}} ^2 - \Gamma_u\!\right)\\
& + \sum_{a=1}^{A}\beta_a\!\left(\frac{1}{4\pi d_a^2} \!\left\Vert \bar{\mathbf{b}}_a^\mathrm{H} ~\mathbf{X} \right\Vert_{\text{F}} ^2- \Gamma_a\!\right),
\end{aligned}
\end{equation}
where $\alpha \in \mathbb{R}_+$, $\boldsymbol{\gamma} \in \mathbb{R}_+^{U \times 1}$, and $\boldsymbol{\beta}\in \mathbb{R}_+^{A \times 1}$ are the Lagrangian multipliers associated with the problem constraints \eqref{powConst1}, \eqref{UExp1}, and \eqref{SExp1}, respectively.
Given the convexity of the Lagrangian function \ac{w.r.t} to $\mathbf{X}$, the optimal solution is found by solving the following system of equations
\begin{align}
\begin{cases}
\frac{\partial \mathcal{L}(\mathbf{X}, \alpha, \boldsymbol{\gamma})}{\partial \mathbf{X}} = 0, \\
\alpha ~\left(\|\mathbf{X}\|_{\text{F}}^2 - P_{\mathrm{max}}\right) = 0,& \alpha \geq 0, \\
\gamma_u \left(\frac{1}{4\pi d_u^2} \|\bar{\mathbf{a}}_u^\mathrm{H}~\mathbf{X}\|_{\text{F}}^2 - \Gamma_u \right) = 0, & \gamma_u \geq 0, \quad \forall u \in \mathcal{U}, \\
\beta_a \left(\frac{1}{4\pi d_a^2} \|\bar{\mathbf{b}}_a^\mathrm{H}~ \mathbf{X}\|_{\text{F}}^2 - \Gamma_a \right) = 0, & \beta_a \geq 0, \quad \forall a \in \mathcal{A}.
\end{cases}
\end{align}
Taking the derivative of $\mathcal{L}$ \ac{w.r.t} $\mathbf{X}$, we obtain
\begin{align}
     \left(\mathbf{A} +2 \alpha~ \mathbf{I}_N +2 \!\sum_{u=1}^{U} \gamma_u \mathbf{C}_u +2 \!\sum_{a=1}^{A} \beta_a \mathbf{E}_a\right)\mathbf{X}\!+2\mathbf{B}=0,
\end{align}
where $\mathbf{C}_u = \frac{1}{4\pi d_u^2}~ \bar{\mathbf{a}}_u \bar{\mathbf{a}}_u^{\mathrm{H}}$ and $ \mathbf{E}_a = \frac{1}{4\pi d_a^2} \bar{\mathbf{b}}_a \bar{\mathbf{b}}_a^{\mathrm{H}}$.

 Consequently, the expression of the optimal waveform is defined as 
\begin{equation}
    \mathbf{X}^\ast = -2 \left( \mathbf{A} + 2\alpha^\ast \mathbf{I}_N + 2\sum_{u=1}^{U} \gamma_u^\ast \mathbf{C}_u  + 2\sum_{a=1}^{A} \beta_s^\ast \mathbf{E}_a\right)^{-1}\! \mathbf{B},
\end{equation}
where $\boldsymbol{\alpha}^\ast$, $\boldsymbol{\gamma}^\ast$ and $\boldsymbol{\beta}^\ast$
are the Lagrangian multipliers obtained by solving the \ac{KKT} equations.

\section{Numerical Results}
\label{Sec5}
In this section, we present the numerical results to validate the proposed design. Throughout all simulations, we set $N=16$ transmit antennas with half-wavelength spacing, $U=4$, $L=30$, $P_{\mathrm{max}}= 12 ~\mathrm{MW}$, and an operating frequency of $6$ GHz. We consider three targets of interest, each positioned at $[-60\degree,0\degree,60\degree]$. Without loss of generality, we assume fixed locations for the targets throughout the simulation while users are distributed randomly.
In our numerical results, we consider one sensitive sector, i.e., $A= 1$, $d_1=300$~m, $\theta_{\text{I}_1}=-28 \degree$, and $\theta_{\text{F}_1}=-11\degree$. We examine the currently established power density restrictions, set at $\Gamma_u =\Gamma = 10 ~\text{W~m}^{-2}, ~\forall u \in \mathcal{U}$, as outlined by both the \ac{FCC} \cite{fields1997evaluating} and the \ac{ICNIRP} \cite{icnirp2020}. However, for the exposure limit of the sector, we choose
$\Gamma_1 =7~\text{W~m}^{-2}$, which is lower than the limit set for general users. This stricter exposure limit is necessary, as sensitive areas require additional protection. These areas include population groups that are more susceptible to \ac{EM} radiations, such as patients and children. 

We evaluate the performance of the proposed design that incorporates \ac{EMF} considerations in comparison to the traditional \ac{RadCom} approach, which optimizes the trade-off without imposing the \ac{EMF} restriction. 
\begin{figure}[t]
    \centering \pgfplotsset{
    every axis/.append style={
        font=\footnotesize,
        legend style={font=\footnotesize,
        legend cell align=left
    },
    compat=1.18
}}
\begin{tikzpicture}
    \begin{axis}[
        xlabel near ticks,
        ylabel near ticks,
        xtick={-90,-60,-30,0,30,60,90},
        xtick pos=bottom,
        ytick pos=left,
        legend pos=north east,
        xlabel={$\theta$ (deg)},
        ylabel={Beampattern (dBi)},
        ymax=7,
        xmax=92,
        xmin=-90,
        ymin=-56,
        width=0.9\linewidth,
        legend entries={Desired beampattern, EMF-aware RadCom,Non EMF-aware RadCom},
        legend pos=south west,
    ]
  
    \addplot[cyan] table {Fig/Beam/RadcomTradeoff.dat};
    \addplot[black!30!green] table {Fig/Beam/dirTradeWEMF.dat};
    \addplot[red] table {Fig/Beam/dir.dat};

        \draw[black,dashed] (axis cs:-28, 20) -- (axis cs:-28, -55);
        \node[align=left, anchor=west, text=black, font=\footnotesize] at (axis cs:-30.1, 5) {$\theta_{\text{I}_1}$};

      \draw[black,dashed] (axis cs:-10.7, 20) -- (axis cs:-10.7, -55);
      \node[align=left, anchor=west, text=black, font=\footnotesize] at (axis cs:-10.5, 5) {$\theta_{\text{F}_1}$};

      \draw[black,dashed] (axis cs:-37.5, 20) -- (axis cs:-37.5, -55);
        \node[align=left, anchor=west, text=black, font=\footnotesize] at (axis cs:-39.3, 5) {$\theta_{1}$};

        \draw[black,dashed] (axis cs:27.1, 20) -- (axis cs:27.1, -55);
        \node[align=left, anchor=west, text=black, font=\footnotesize] at (axis cs:27.2, 5) {$\theta_2$};

        \draw[black,dashed] (axis cs:11.7, 20) -- (axis cs:11.7, -55);
        \node[align=left, anchor=west, text=black, font=\footnotesize] at (axis cs:11.8, 5) {$\theta_3$};
 
    \draw[black,dashed] (axis cs:80.9, 20) -- (axis cs:80.9, -55);
        \node[align=left, anchor=west, text=black, font=\footnotesize] at (axis cs:80.1, 5) {$\theta_4$};
    \end{axis}
\end{tikzpicture}
    \caption{Radar beampatterns considering different approaches for $\rho=0.2$.}
    \label{fig:beam}
\end{figure}
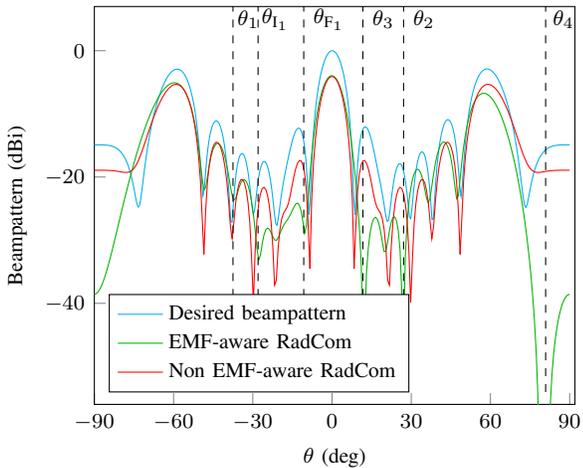  

Initially, we present an assessment of radar performance, focusing on the beampattern. In Fig.~\ref{fig:beam}, we depict the users' locations in terms of angles along with the beampattern for three approaches; (\textit{i}) \ac{RadCom} without considering the \ac{EMF} constraints, (\textit{ii}) the proposed \ac{EMF}-aware \ac{RadCom}, and (\textit{iii}) the desired beampattern obtained by solving the \ac{MUI} minimization problem \eqref{X0}. 
As shown in Fig.~\ref{fig:beam}, the proposed \ac{EMF}-aware design aims to minimize the power of the beam directed towards users' locations and sensitive sectors. This design tends to preserve the main beam's shape towards the radar targets while reducing the sidelobes directed towards users and sensitive areas. In simple terms, this can be seen as shifting sidelobes away from angles where exposure minimization is desired. Notably, users' locations can be conceptualized as sectors characterized by small angular domains.
\begin{figure}[t]
    \centering
    \pgfplotsset{
    every axis/.append style={
        font=\footnotesize,
        legend style={font=\footnotesize,
        legend cell align=left
    },
    compat=1.18
}}

    \begin{tikzpicture}
        \begin{semilogyaxis}
            [xlabel near ticks,
            ylabel near ticks,
             xtick pos=bottom,
            ytick pos=left,
            legend pos=north east,
            ymajorgrids=true,
            xmajorgrids=true,
            grid style=dashed,
            xlabel={$\mathrm{SNR}$ (dB)},
            ylabel={$\mathrm{SER}$},
            xmax=25,
            ymin = 10e-6,
            ymax = 1,
            grid = both,
            width=0.9\linewidth,
            legend entries={Non EMF-aware RadCom, EMF-aware RadCom $\Gamma= 10~\text{W~m}^{-2}$, EMF-aware RadCom $\Gamma= 0.1~\text{W~m}^{-2}$},
            legend pos=south west]
            \addplot[red,mark=square,
            line width=1pt] table {Fig/SER_Fig/SERErDir.dat};
            \addplot[black!30!green,mark=triangle,
            line width=1pt] table {Fig/SER_Fig/SERErDirWEMF10.dat};
            \addplot[blue,mark=o,
            line width=1pt] table {Fig/SER_Fig/SERErDirWEMF1.dat};

                      \end{semilogyaxis}
    \end{tikzpicture}
\caption{Symbol error rate, $\mathrm{SER}$, vs. $\mathrm{SNR}$ for different \ac{EMF} thresholds $\Gamma$ for $\rho=0.4$.}
\label{fig:SER} 
\end{figure}
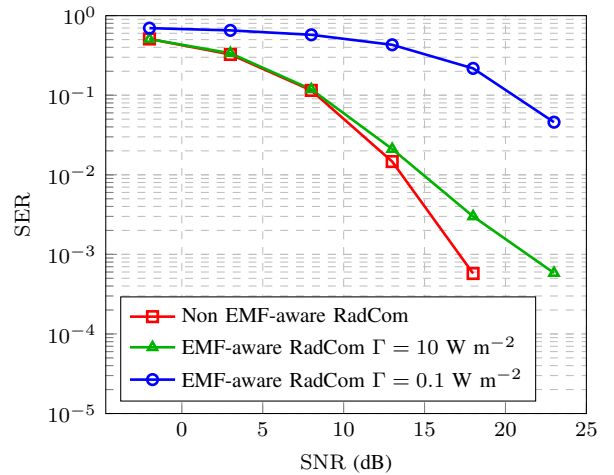
 
In Fig.~\ref{fig:SER}, we show the performance of the proposed \ac{RadCom} system in terms of \ac{SER} as a function of the transmit \ac{SNR} for different power density thresholds for the \ac{EMF} constraints, where the \ac{SNR} is defined as $\mathrm{SNR} = P/\sigma_0^2$. The results show that our solution performs comparably to the directional beampattern without \ac{EMF} constraints that points to the directions of interest, with only slight performance loss in the radar beampattern. On the contrary, it yields a substantial reduction in downlink exposure. For instance, the total users' exposure in a \ac{RadCom} scenario without \ac{EMF} exposure constraints equals to $\overline{\mathrm{S}} = \sum_{u=1}^U \mathrm{S}_{\text{U}}(\theta_u,d_u)=71,63~\text{W~m}^{-2}$. However, when using \ac{EMF}-aware design, the total exposur is reduced to $40 ~\text{W~m}^{-2}$ while maintaining nearly the same \ac{SER} for an $\mathrm{SNR}$ of $3$ dB. Additionally, when the threshold is stringent, the system may struggle to meet \ac{RadCom} performance requirements. 
 
 Similarly, in Fig.~\ref{fig:SERvsRho}, the designed radar waveform only experiences slight performance loss compared to the \ac{RadCom} waveform without considering the \ac{EMF} constraint. For a fixed $\mathrm{SNR}= 13$~dB, the $\mathrm{SER}$ decreases significantly by increasing the trade-off factor $\rho$. As $\rho$ tends towards unity, the system only minimizes the \ac{MUI}, prioritizing the communication.
\begin{figure}[t]
    \centering
    \pgfplotsset{
    every axis/.append style={
        font=\footnotesize,
        legend style={font=\footnotesize,
        legend cell align=left
    },
    compat=1.18
}}

\begin{tikzpicture}
    \begin{semilogyaxis}[
        xlabel near ticks,
        ylabel near ticks,
         xtick pos=bottom,
        ytick pos=left,
        ymajorgrids=true,
        xmajorgrids=true,
        grid style=dashed,
        xlabel={$\rho$},
        ylabel={$\mathrm{SER}$},
        xmin=0,
        xmax  =1,
        ymax = 7,
        grid = both,
        width=0.9\linewidth,
        legend entries={Non EMF-aware RadCom, EMF-aware RadCom $\Gamma= 10~\text{W~m}^{-2}$,EMF-aware RadCom $\Gamma= 0.1~\text{W~m}^{-2}$},
        legend pos= north west,
    ]
    \addplot[red, mark=square,
            line width=1pt] table {Fig/SER_Fig_Rho/SERRhoOpt10.dat};
     \addplot[black!30!green, mark=triangle,
            line width=1pt] table {Fig/SER_Fig_Rho/SERRhorDir1.dat};
    \addplot[blue, mark=o,
            line width=1pt] table {Fig/SER_Fig_Rho/SERRhoOpt01.dat};
       
    \end{semilogyaxis}
\end{tikzpicture}
\caption{Symbol error rate, $\mathrm{SER}$, vs. trade-off factor $\rho$ for different \ac{EMF} thresholds $\Gamma$.}
\label{fig:SERvsRho} 
\end{figure}
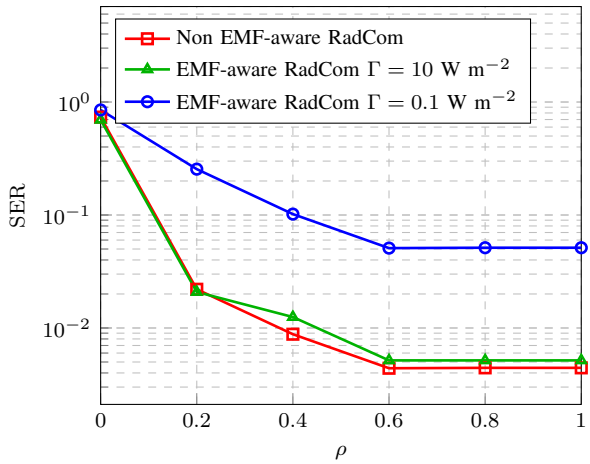
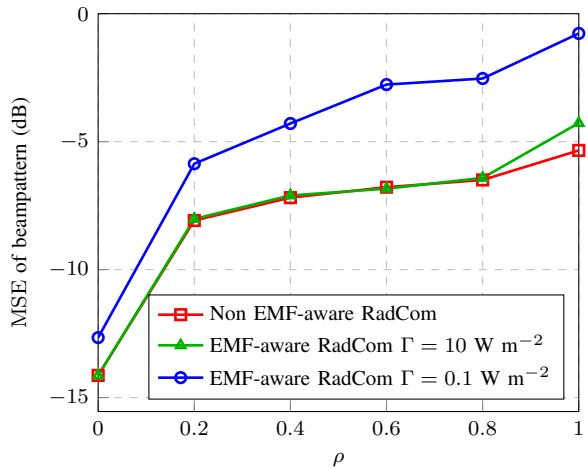
\begin{figure}[t]
    \centering
    \pgfplotsset{
    every axis/.append style={
        font=\footnotesize,
        legend style={
            font=\footnotesize,
            legend cell align=left,
        },
        compat=1.18,
    },
}
\begin{tikzpicture}
    \begin{axis}[
        xlabel near ticks,
        ylabel near ticks,
         xtick pos=bottom,
        ytick pos=left,
        xtick={0,0.2,0.4,0.6,0.8,1},
        xmin=0,
        xmax=1,
        ymax = 0,
        ymajorgrids=true,
        xmajorgrids=true,
        grid style=dashed,
        xlabel={$\rho$},
        ylabel={$\mathrm{MSE}$ of beampattern (dB)},
        grid =both,
        width=0.9\linewidth,
        legend entries={Non EMF-aware RadCom, EMF-aware RadCom $\Gamma= 10~\text{W~m}^{-2}$,EMF-aware RadCom $\Gamma= 0.1~\text{W~m}^{-2}$},
        legend pos=south east,
    ]
        \addplot [red,mark=square,
            line width=1pt] table {Fig/SquaredErr_Fig/BER/DirMSE.dat};
           \addplot [black!30!green,mark=triangle,
            line width=1pt] table {Fig/SquaredErr_Fig/BER/Opt10MSE.dat};
        \addplot [blue,mark=o,
            line width=1pt] table {Fig/SquaredErr_Fig/BER/Opt01MSE.dat};
    \end{axis}
\end{tikzpicture}
\caption{Mean Squared error between the desired and the Radcom beampattern $\mathrm{MSE}$ vs. trade-off factor $\rho$ for different \ac{EMF} thresholds $\Gamma$.}
\label{fig:MSE}
\end{figure}

In Fig.~\ref{fig:MSE}, we investigate the radar performance in terms of the mean squared error $\mathrm{MSE}$ between the desired beampattern and the designed one. The radar's overall performance remains largely unaffected when adhering to the conventional threshold set by \ac{ICNIRP}. Given the inherent trade-off between the system's two performances, the \ac{EMF} constraint affects the communication aspect of the system more severely than the radar performance, especially for strict \ac{EMF} conditions.


\section{Conclusion}
\label{Sec6}
In this work, we investigate the waveform design with the \ac{EMF} constraints in terms of the incident power density in downlink transmission for multi-user \ac{MIMO} dual-functional \ac{RadCom} system. Then, we formulate a convex optimization problem to obtain an \ac{EMF}-aware waveform. Finally, our numerical results show that the proposed design achieves up to 40\% \ac{EMF} exposure reduction by redirecting the sidelobes of the radar waveform away from the users' locations and away from the restricted sectors.

\enlargethispage{0cm}
\bibliographystyle{IEEEtran}
\enlargethispage{0cm}
\bibliography{EMF-awareRadcom}
\end{document}